# Nonlinear X-ray Compton Scattering

Matthias Fuchs[1,2], Mariano Trigo[2,3], Jian Chen[2,3], Shambhu Ghimire[2,3], Sharon Shwartz[4], Michael Kozina[2,3], Mason Jiang[2,3], Thomas Henighan[2,3], Crystal Bray[2,3], Georges Ndabashimiye[2], Philip H. Bucksbaum[2,3,7], Yiping Feng[5], Sven Herrmann[6], Gabriella Carini[6], Jack Pines[6], Philip Hart[6], Christopher Kenney[6], Serge Guillet[5], Sébastien Boutet[5], Garth J. Williams[5], Marc Messerschmidt[5,8], M. Marvin Seibert[5], Stefan Moeller[5], Jerome B. Hastings[5], David A. Reis[2,3,7]

[1]*Department of Physics and Astronomy, University of Nebraska - Lincoln, Lincoln, NE 68588, USA*

[2]*Stanford PULSE Institute, SLAC National Accelerator Laboratory, Menlo Park, CA 94025, USA*

[3]*Stanford Institute for Materials and Energy Sciences, SLAC National Accelerator Laboratory, Menlo Park, CA 94025, USA*

[4]*Physics Department and Institute of Nanotechnology, Bar Ilan University, Ramat Gan, 52900, Israel*

[5]*Linac Coherent Light Source, SLAC National Accelerator Laboratory, 2575 Sand Hill Road, Menlo Park, California 94025, USA*

[6]*SLAC National Accelerator Laboratory, Menlo Park, CA 94025, USA*

[7]*Department of Photon Science and Applied Physics, Stanford University, Stanford, CA 94305, USA*

[8]*BioXFEL NSF Science and Technology Center, 700 Ellicott Str., Buffalo, NY 14203, USA*

**X-ray scattering is typically used as a weak linear probe of matter. It is primarily sensitive to the spatial position of electrons and their momentum distribution[1,2]. Elastic X-ray scattering forms the basis of atomic-scale structural determination[3]**

while inelastic Compton scattering[4] is often used as a spectroscopic probe of both single-particle excitations and collective modes[5]. X-ray free-electron lasers (XFELs) are extraordinary tools for studying matter on its natural time and length scales due to their bright and coherent ultrashort X-ray pulses. However, in the focus of an XFEL the assumption of a weak linear probe breaks down, and nonlinear light-matter interactions can become ubiquitous[6-15]. The electromagnetic field can be sufficiently high that even non-resonant multiphoton interactions at hard X-ray wavelengths become relevant. Here we report the observation of one of the most fundamental nonlinear X-ray–matter interactions, the concerted Compton scattering of two identical photons producing a single higher-energy photon. We measure scattered photons with an energy near 18 keV generated from solid beryllium irradiated by XFEL pulses near 9 keV. The intensity in the X-ray focus reaches up to $4 \times 10^{20}$ Watt/cm$^2$, which corresponds to a peak electric field two orders of magnitude higher than the atomic unit of field-strength and within four orders of magnitude of the quantum electrodynamic critical field[16,17]. The observed signal is well above background. It scales quadratically in intensity and is emitted into a non-dipolar pattern, consistent with the simultaneous two-photon scattering from a collection of free electrons. However, the energy of the generated photons shows an anomalously large redshift only present at high intensities. This indicates that the instantaneous high-intensity scattering effectively interacts with a different electron momentum distribution than linear Compton scattering, with implications for the study of atomic-scale structure and dynamics of matter.

In linear Compton scattering a photon transfers energy and momentum to an electron during the scattering process. The spectrum of scattered photons at a given momentum transfer is a direct probe of a material's electron momentum distribution when the energy loss is large compared to the relevant binding energies (the impulse approximation)[18]. At the focus of an X-ray free-electron laser, nonlinear X-ray–matter interactions can become important. For elastic X-ray scattering in crystals the second order nonlinearity has been considered theoretically as originating from the classical anharmonic motion of a periodic collection of free electrons[19,20]. For free electrons, nonlinear scattering was described semiclassically within the framework of quantum electrodynamics (QED) by Brown and Kibble half a century ago[21]. The dynamics of the interaction and the scattering rate depend strongly on the normalized vector potential $\eta = eE/m_e c\omega$, which represents the root-mean-square (rms) transverse momentum imparted to the electron by a classical electromagnetic wave of rms field-strength $E$ and angular frequency $\omega$; here $e$ is the elementary charge, $m_e$ the electron rest-mass and $c$ the speed of light. Free-electron models have been successful at describing nonlinear X-ray matter interactions, including phase-matched X-ray–optical sum frequency generation[22] and X-ray second harmonic generation[15]. Here we report the observation of nonlinear X-ray Compton scattering from solid beryllium. Our measurements significantly differ from free-electron results, suggesting an increased role of bound-state electrons in this nonlinear interaction.

Previous experiments have investigated the interactions of electrons with high-intensity (~$10^{18}$ W/cm$^2$) optical radiation in the relativistic-regime, $\eta \sim 1$, i.e. when the electron gains an energy on order of its rest mass when accelerated across a distance equal to the radiation wavelength. They include the generation of harmonics from plasma electrons[23] and multiphoton Compton scattering from a beam of ultrarelativistic free electrons in a near head-on collision geometry[24].

To approach this relativistic regime with hard X-rays requires intensities ~$10^{26}$ W/cm$^2$, well beyond what is currently achievable with XFELs. Nonetheless, concerted two-photon scattering processes should be observable from solid targets, at orders of magnitude less intensity, based on perturbative scaling. For $\eta \ll 1$, a free electron will undergo anharmonic motion with the $n^{th}$ harmonic contribution to the induced current $\sim \eta^n$, and thus the cross-section for nonlinear scatter will scale as $\eta^{2n-2} r_0^2$, where $r_0$ is the classical electron radius. The efficiency in the case of (elastic) phase-matched X-ray second harmonic generation was measured to be $6 \times 10^{-11}$ for peak fields of $\sim 10^{16}$ W/cm$^2$ within an angular acceptance of 180 μrad—in agreement with the free-electron model[15].

Here we report the observation of the concerted nonlinear Compton scattering of two hard X-ray photons with variable energy around 9 keV in beryllium producing a single ~18 keV photon. The maximum intensity was ~$4 \times 10^{20}$ W/cm$^2$ corresponding to a peak electric field ~$5 \times 10^{11}$ V/cm ($\eta \sim 2 \times 10^{-3}$), well in the perturbative regime. We observe a signal that varies quadratically with the FEL intensity and is well above the measured background. It is emitted in a non-dipolar pattern as expected for a second order perturbative process. However, the photon spectrum shows an anomalously large redshift in the nonlinearly generated radiation compared to both QED calculations for free electrons initially at rest and to the simultaneously measured linear scattering from the low-intensity residual FEL second harmonic generated by the undulators[25]. Since the Compton redshift is a consequence of momentum conservation in the electron-photon system, this anomalous redshift suggests that the two-photon scattering cross-section has contributions from electron momenta than are significantly different than linear scattering from the ground-state distribution. We chose beryllium because it is a low Z material with relatively low photoionization cross-section, favourable ratio of Compton to elastic linear cross-sections[26] as well as high melting point. In addition the linear Compton profile in beryllium has been measured with extremely high resolution out to more than 1.5 atomic units using similar x-ray photon energies[27]. In those studies, details of the conduction band structure as well as a contribution from the 1$s$ core electrons are evident.

Importantly the linear scattering from Be is well described in the free-electron, impulse approximation.

The strong X-ray fields were produced in the ~100 nm focus of the linearly polarized, ~1.5 mJ, 50 fs, Linac Coherent Light Source XFEL, using the Coherent X-ray Imaging (CXI) instrument[28]. The incident X-ray energy was tuned in the range of 8.8–9.75 keV. Two solid Be targets were arranged at a 45º incident angle (see figure 1), one at the X-ray focus and the other downstream of the focus where the intensity is low. The effective target thickness (1.4 mm) is considerably smaller than the linear absorption length of the fundamental (7.3 mm for 9 keV), such that > 80% of the beam is transmitted. The angular distribution of the scattered radiation was detected using multiple 2D pixel array detectors[29] arranged in an arc with radius ~20 cm covering observation angles from 80º – 135º in the polarization plane of the XFEL (see figure 1). The detectors have an intrinsic energy resolution that allows us to distinguish between a single ~9 keV and ~18 keV photon, but cannot differentiate between a single ~18 keV photon and the pile-up of two ~9 keV photons absorbed in a single pixel during a single shot. In order to mitigate the background due to pile-up, the linear scattering from the FEL fundamental was sufficiently attenuated by placing 250 μm thick Zr foils directly in front of the detectors. This largely transmits (~10%) X-rays just under the Zr K-edge ($\omega_K = 17.996$ keV), while attenuating photons around 9 keV and just above 18 keV by about seven orders of magnitude (see Methods). The overall background was simultaneously characterized with a nearly identical configuration placed well out of focus, at a significantly lower FEL intensity but with a comparable number of incident photons (see figure 1). Additional background originates from the linear scattering of the FEL second harmonic that is produced in the undulators[25]. This is substantially reduced before reaching the target by the low reflectivity of the focussing mirrors for photon energies above 11 keV[28]. The scattering of the residual second FEL harmonic was measured to be negligible in comparison to the nonlinear signal; no significant scattered third FEL harmonic was detected.

Consider the concerted scattering of *n*-photons each with energy $\omega$ and momentum $\boldsymbol{k}$ from a free electron of initial momentum $\boldsymbol{p}$, into a single photon ($\omega_n'$, $\boldsymbol{k_n}'$). Energy and momentum conservation lead to the generation of a single photon with energy

$$\omega_n'(\omega, \theta) = \frac{n\omega + \frac{\boldsymbol{p} \cdot \boldsymbol{K}_n}{m_e}}{1 + \left(\frac{n\omega}{m_e} + \frac{\eta^2}{2}\right)(1 - \cos\theta)}, \qquad (1)$$

where $\boldsymbol{K}_n = n\boldsymbol{k} - \boldsymbol{k_n}'$ is the *n*-photon momentum transfer, $\cos\theta = \boldsymbol{k} \cdot \boldsymbol{k_n}'/\omega\omega_n'$, and we have set $\hbar = c = 1$. The expected two-photon spectrum $\omega_2'(\omega_0 = 9\ keV, \theta = 90°)$ will be roughly centred around 17.4 keV, i.e. redshifted by approximately 0.6 keV from the 2$^{nd}$ harmonic at 18 keV, such that the redshift $\Delta_2 = 2\omega_0 - \omega_2'$ is well above the 1*s* binding energy of Be (0.112 keV). For our parameters, ($\eta \sim 2\times10^{-3}$), ponderomotive effects are expected to contribute negligibly to the kinematics as the ponderomotive energy, $\eta^2 m_e/2 = 0.6$ eV $\ll \omega_0$. Thus, the kinematics for *n*-photon nonlinear scattering from a free electron is to very high degree the same as for the linear (*n*=1) scattering of a single photon with *n* times the energy, i.e. $\omega_1'(\omega = n\omega_0, \theta)$ (such as from the harmonics produced in the undulator by the FEL process). Nonetheless, nonlinear *n*-photon processes can be distinguished from the linear case as the rate and angular distribution of nonlinear scattering depend strongly on both the radiation field strength and the order $n$ of the process[30]. As the nonlinear cross-section in the perturbative regime scales as $\sigma^{(n)} \sim \eta^{2n-2}$ the scattering signal scales with the incoming intensity as $I^n$ (quadratically for an *n*=2 process). In addition, the differential cross-section shows a distinctly different angular emission than the Klein-Nishina formula[31] for linear scattering: for the *n*=2 perturbative process, the photons scattered from free electrons initially at rest are expected to be emitted in an asymmetric quadrupole-like pattern peaked at an angle of ~130º (in the backward direction, see figure 1b)[21,30]. Importantly, it has a finite emission along the FEL polarization (90º) where the linear scattering (dipole emission, fig. 1c) is strongly suppressed.

We measure the dependence of the signal on the FEL fundamental intensity at $\omega_0 = 9.25$ keV by inserting silicon attenuators of different thicknesses into the beam before the focusing mirrors. The observed signal at a photon energy near $\omega' \sim 18$ keV from the high-field interaction region (fig. 2a top row) shows a significant dependence on the FEL pulse energy that is not present in the low-intensity signal (fig. 2a bottom row). In addition there is substantial scattering at $\omega' \sim 18$ keV in the ~90º observation direction for the high-field interaction. This in combination with the quadratic pulse-energy dependence (fig 2b) is consistent with an *n*=2 nonlinear scattering process. For the low-intensity interaction region the signal is nearly independent of the incident FEL fundamental pulse energy and there is relatively low scattering near 90º. This is consistent with *n*=1 linear scattering from the residual FEL second harmonic, i.e. $\omega'_1(2\omega_0, \theta)$, considering the nearly constant transmission of ~18 keV photons through the Si attenuators. As further evidence of a nonlinear interaction, the *n*=2 signal shows strong dependence on the positioning of the sample relative to the focal plane as shown in Figure 3 (measured at the maximum incident pulse energy, and at $\omega_0 = 8.8$ keV). The *n*=2 signal is strongly reduced when the sample is translated on order of its thickness through the focus, confirming that the dominant nonlinear scattering occurs from the high-intensity region close to the focal plane.

We explore the spectrum of the *n*=2 scattered photons using the strong variation in transmittance of the filter around the Zr K-edge ($\omega_K$) as a coarse spectrometer (see Figure 4 and Methods). By varying the incident photon energy, this allows us sensitivity to scattered photons with redshift, $\Delta_2 > 2\omega_0 - \omega_K$. According to equation (1), $\omega'_2$ is centered about $\omega_K$ for $\omega_0 = 9.28$ (9.59) keV at the minimum(maximum) detection angle of 82° (138°). The low-intensity signal $\omega'_1(2\omega_0, \theta) \approx \omega_K$ (Figure 4 b) shows a strong energy and angle dependence with a cutoff that is consistent with linear Compton scattering of the 2$^{nd}$ FEL harmonic from the loosely bound electrons in Be. However, the nonlinear signal $\omega'_2(\omega_0, \theta) \approx \omega_K$ at high intensity (Figure 4 a) shows no evidence of a cutoff even up to the highest incident fundamental photon energy. This corresponds to a total energy loss $\Delta_2 > 1.5$ keV at $\omega_0 = 9.75$ keV. At $\theta = 90°$ the predicted Compton shift for the $n = 2$ process is centered ~700 eV, and thus there is substantial nonlinear scattering at an additional redshift of at least 800 eV that is absent in the linear signal.

The observation of scattered photons with an additional redshift of this magnitude requires that significant momentum must be supplied either by the medium in the initial state or taken up by the medium in the final state. The minimum required momentum transfer is comparable to the typical momentum of a 1s electron of Be ($p_a = Z/a_0 = 15$ keV/c, where $a_0 \sim 0.5$ Å is the Bohr radius) but considerably less than the momentum of a primary photoelectron (~100 keV/c). This suggests either a process with preferential nonlinear scattering from bound electrons, with the missing momentum carried in the final state by the recoil of the ion, or nonlinear scattering from secondary plasma electrons following photoionization within the same FEL pulse.

We note that there is a weak bound-state contribution to the linear Compton scattering background from the second FEL harmonic. The predominance of the large redshift *n*=2 scattering, however, suggests that any bound-state contribution must involve breakdown of the free-electron approximation from the ground-state momentum distribution and a new scattering mechanism. The interaction term in the light-matter

Hamiltonian is given by $H_{int} = \frac{e}{m} \mathbf{A} \cdot \mathbf{p} + \frac{e^2 A^2}{2m}$, where $\mathbf{A}$ is the vector potential operator and $\mathbf{p}$ the electron momentum operator. For a given final state, and in second order in $H_{int}$, there will be quadratic processes in the intensity involving: an intermediate state comprising (1) one photon being annihilated from the field, a free-electron and ion and (2) a single photon annihilated from the field, the creation of a scattered photon, and a recoiled atom (in the ground electronic state). When the momentum transfer is similar to the typical 1s electron momentum, process (1) is expected to dominate. In this case, we estimate that the total atomic cross-section could be comparable to that of a (hot) free electron, although the angular distributions will differ. For a larger momentum transfer, red-shifts comparable to the photon energy are possible at much higher rates, but these would not be easily observable in the current experiment due the restricted energy resolution and strong frequency dependent attenuation from Zr filters.

Alternatively for the missing momentum to be supplied by an initial hot electron, it needs to have a component of its kinetic energy along $\mathbf{K_2}$, $\frac{\hbar^2 q_{\parallel}^2}{2m} > 156$ eV. While we cannot completely rule out this process, we note that for a thermal distribution, this would to lowest order produce a Gaussian broadened, but not shifted spectrum with half-width, half-maximum of $\sqrt{2ln(2)kT/m}\, K_2$ requiring a plasma temperature $kT$ ~320eV (3.7 ×10$^6$ K) to broaden to 18 keV for $\omega_0$ =9.75 keV and $\theta$ =90º. For comparison we expect $kT$ ~110 eV at the highest intensity for photo-ionization cross-section of 6.9 barns/atom assuming a fully ionized and thermalized plasma develops within the focal volume after the pulse.

Thus additional experiments are required to distinguish the relative contributions from nonlinear scattering from the nonequilibrium plasma and the bound-state contributions. We note that process (2) above can be phase-matched for a crystalline sample when the single photon momentum transfer, $K_1$ equals a reciprocal lattice vector. This suggests that experiments on single crystals, could yield much higher nonlinearities, and thus could have implications for structure determination at high intensity.

**Acknowledgements**

This work was supported primarily by the US Department of Energy (DOE), Office of Basic Energy Sciences (BES) and the Volkswagen Foundation. Portions of this research were carried out at the Linac Coherent Light Source (LCLS) at the SLAC National Accelerator Laboratory. Preparatory measurements were carried out at the Stanford Synchrotron Radiation Lightsource (SSRL). Both, LCLS and SSRL are Office of Science User Facilities operated for the U.S. Department of Energy Office of Science by Stanford University. M.F. acknowledges support from the Volkswagen Foundation. M.K. was supported by the DOE Office of Science Graduate Fellowship Program. M.T. and J.C. were supported by the Division of Materials Sciences and Engineering, BES, DOE under contract 51 DE-AC02-76SF00515. D.A.R. G.N. and S.G. were supported by the AMOS program within the Chemical Sciences, Geosciences, and Biosciences Division, DOE, BES, DOE. We thank Dr. R. Santra for discussions.

**Author contributions.**

M.F. and D.A.R. conceived and with S.B., C.K., S.Gu. and J.B.H. designed the experiment. S.H., G.C., J.P., P.H., C.K., S.Gu., S.B., G.J.W. and M.M designed and fabricated components of the experiment. M.F., M.T., J.C., S.Gh., S.S., M.K., M.J., T.H., C.B., G.N., Y.F., S.H., S.M, J.B.H. and D.A.R. carried out the experiment. S.B., G.J.W., M.M. and M.M.S. operated the coherent X-ray imaging instrument. M.F., M.T. and J.C. analysed the data. M.F and D.A.R interpreted the results with input from P.H.B, and M.F. and D.A.R. wrote the manuscript with input from all other authors.

**Author Information**

The authors declare no competing financial interests.

Correspondence and requests for materials should be addressed to: mfuchs@unl.edu

**Methods**

X-ray pulses with a photon energy tuned from 8.8 – 9.75 keV and ~0.5% FWHM bandwidth were focused to a nominally ~100 nm spot using Kirkpatrick-Baez focusing mirrors (depth of focus: ~0.2 mm). The pulses had an energy of up to ~1.5 mJ on target and an averaged pulse duration of ~50 fs (envelope), which leads to a focused peak intensity of $4 \times 10^{20}$ W/cm$^2$ ($\eta \sim 2 \times 10^{-3}$ for 9 keV). The precise pulse duration and focal size could not be measured. A 1 mm thick polycrystalline piece of high-purity solid beryllium (Materion PF-60, with >99% beryllium content and a low level of heavy impurities) oriented at an angle of 45° was used as target. Despite the small X-ray cross-sections for scattering and photoionization in Be (attenuation length of 7.3 mm for 9 keV photons[32]), the photon flux is sufficiently high that the interaction ultimately leads to plasma formation and irreversible damage in a single shot. The sample was stepped to a fresh spot each shot. Scattered photons were detected each shot at 120 Hz using CSPAD 140K 2D pixel array detectors (PADs)[29] located at observation angles 80 – 135º (see figure 1). The detectors were substantially shielded from scattering of the FEL fundamental by 250 μm thick zirconium filters positioned directly in front of the detector (suppression of the 9 keV signal by approximately seven orders of magnitude while transmitting 10% of the signal just below the Zr K-edge at 17.996 keV). Each detector has 140,000 pixels, which allowed the spatial discrimination of pixels that exceed a particular photon count rate (such as from powder diffraction of the polycrystalline Be). The probability for photon pile-up in the detector at full beam intensity was decreased to $\sim 10^{-8}$ counts/pixel/shot (3 orders of magnitude below the measured signal), by a combination of the Zr filter and by post-detection software masking of any pixel that (on average) measured a rate of fundamental photons exceeding $10^{-4}$ counts/shot. Post software-correction, each detector image shows a slowly varying spatial distribution of near 18 keV photons. The CSPAD detectors had a coarse energy-resolution (few keV), from which the approximate energy deposited by a photon into each pixel could be deduced. We used Cu fluorescence and the scattering signal of the FEL fundamental and 2$^{nd}$ harmonic for the ADU to photon energy conversion in fig. 2. The harmonic content in the FEL beam (generated in the undulator) at hard X-ray energies is < 0.1% (2$^{nd}$ harmonic) and <2% (3$^{rd}$ harmonic) compared to the fundamental power[33]. The FEL harmonics were suppressed before reaching the sample by grazing incidence focusing mirror reflections (low reflectivity for energies above 11 keV), leading to at least seven orders of magnitude fewer photons on target compared to the FEL fundamental. We measured the maximum background from linear scattering of the 2$^{nd}$ FEL harmonic to vary from 4 ×10$^{-8}$ photons/shot/pixel near 90° to 4 ×10$^{-7}$ photons/shot/pixel near 135º and no significant 3$^{rd}$ FEL harmonic.

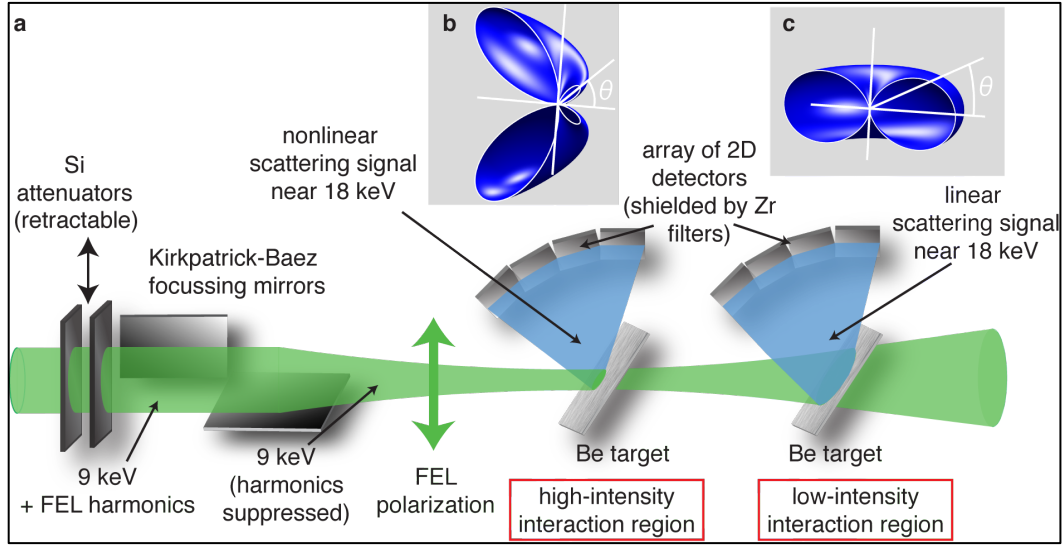

**Figure 1 | Experimental setup. a)** The XFEL beam with photon energy near 9 keV (green) was incident on two 1mm thick Be targets oriented at an incident angle of 45 located in a He environment. A high- and a low-intensity interaction region allowed the simultaneous measurement of the signal and background as the X-rays are transmitted largely unattenuated by the Be. The first target was placed in the ~100 nm XFEL focus (high intensity interaction region), and a second identical target was placed well out of focus (low intensity interaction region). The average pulse energy of the FEL fundamental was varied using Si attenuators. The weaker residual FEL second harmonic and third harmonic (generated in the undulator) were significantly rejected by the focusing mirrors before reaching the target. Scattered radiation from the two interaction regions was collected over a wide solid angle using arrays of four 2D detectors (each detector comprising 140k pixels covering ~14°x14° centred at 131°,117°,103° and 89° in the polarization plane for the high intensity arc, which due to experimental constraints was rotated by -5° relative to the low intensity arc). 250 μm thick Zr foils placed directly in front of the detectors significantly attenuated the scattered photons near the FEL fundamental while largely transmitting photons with energies just under the Zr K-edge (17.996 keV). The incident photon energy was chosen so that the Zr filter acts as a coarse

spectrometer for resolving the energy loss in the two-photon Compton signal, $\omega_2'(\omega_0, \theta)$. b) shows a calculation of the expected angular distribution of the free-electron nonlinear second harmonic emission generated by a free electron at rest. In the perturbative regime (vector potential $\eta \sim 2 \times 10^{-3}$), the emission is peaked at a scattering angle of $\theta \sim 130°$ and includes a finite scattering into the FEL polarization direction $\theta = 90°$. c) shows the calculated emission pattern for low-intensity linear scattering (dipole emission), which has a negligible scattering contribution into the polarization direction.

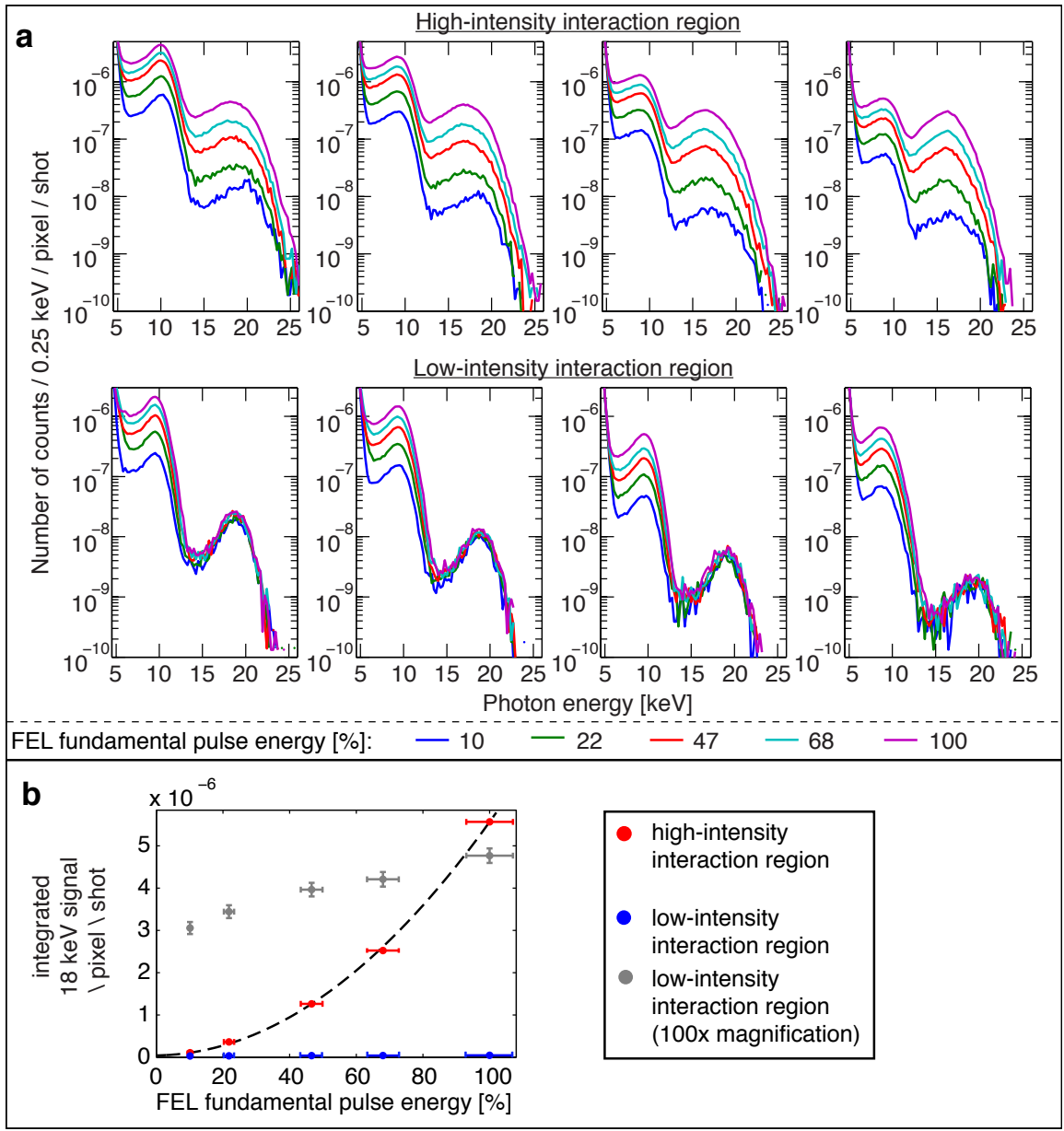

**Figure 2 | Scattering signal as a function of the FEL fundamental pulse energy. a** shows the signals from the high-intensity interaction (top row) and the low-intensity interaction region (bottom row) for FEL fundamental pulse energies ranging from 10-100% of the full beam (colour coded) at an incident photon energy of 9.25 keV. The detectors were centred at observation angles of (from left to right) 131°, 117°, 103°, 89° and each detector recorded a solid angle of ~14º x 14º. Due to experimental constraints, the detectors observing the sample out of focus (bottom row) were rotated by 5° (centred

at 136-94°). Only the high-intensity interaction detectors showed a significant, nonlinear change with the FEL intensity in the near 18 keV signal. In particular at 90º (right column) where the dipole emission pattern of linear scattering is highly suppressed (fig 1c), we measured significantly more photons relative to the low-intensity interaction (bottom row) as well as the FEL fundamental (at 9.25 keV). Each plot shows the sum of ~170,000 histograms which were individually recorded for each FEL shot. They represent the average photon energy deposited in a single detector pixel deduced from the generated charge in a pixel and are normalized by the number of shots. The detected signal is broadened by the electronic noise of the detector. The intensity of the FEL fundamental was varied using silicon absorbers before focusing the beam. The effect of the attenuators on the transmission of the residual FEL second harmonic is comparably small (73% at the highest attenuation). Note that the detector signal on any single shot is sparse as the probability of detecting a photon in a given pixel is negligible. **b** shows the integrated signal near $\omega_2'(\theta)$ ~18 keV as a function of the FEL intensity for the detectors near 90°. Red represents the signal originating from the high-intensity interaction, blue from the low-intensity interaction and grey is the low-intensity signal multiplied by a factor of 100 for magnification. The dotted line shows a quadratic fit as a function of FEL intensity for the high intensity signal as is expected for a second order nonlinear effect. The signal from the low-intensity interaction varies only slightly, consistent with linear scattering of the residual second FEL harmonic and weak attenuation by the Si filters. The vertical error bars for the blue and red data points on this scale are smaller than the marker size.

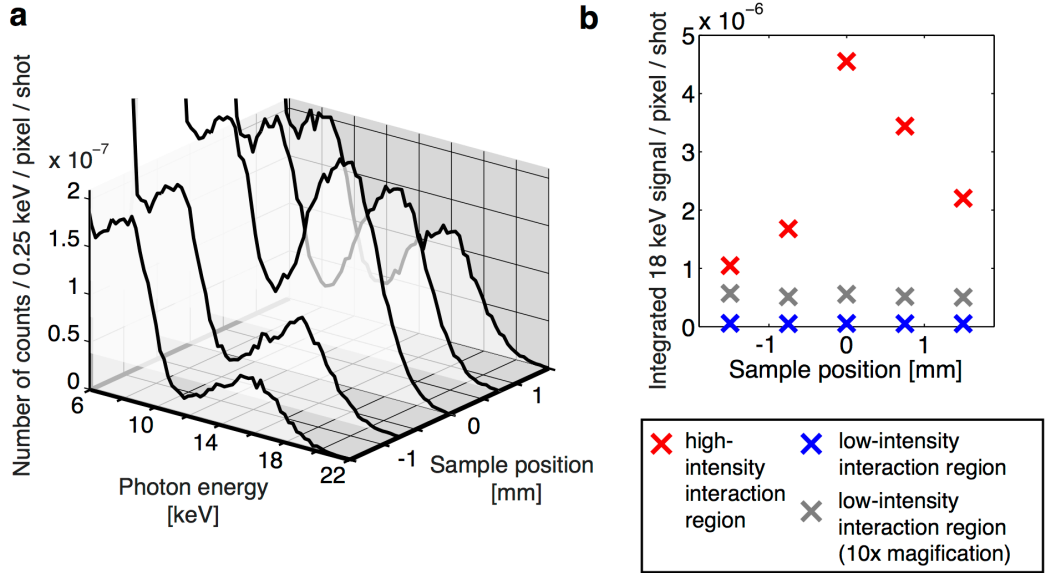

**Figure 3 | Dependence of nonlinear Compton scattering on sample position relative to the FEL focus. a** shows histograms of the detector observing the high-intensity interaction region positioned at an observation angle of 90° for different sample positions with respect to the nominal FEL focal plane. The FEL was unattenuated and the fundamental wavelength for this scan was 8.8 keV. Each histogram is averaged over ~60,000 shots. The effective sample length (at 45º) is 1.4 mm. The ~18 keV signal varies strongly with the sample position whereas the ~9 keV signal from the high-intensity region and the overall low-intensity signal (over the whole energy range) remains practically constant. Note that the detectors at the other observation angles show a similar behaviour. **b** shows the signal integrated signal around the 18 keV peak versus the sample position for the high-intensity interaction (red) compared to the peak at the low-intensity (fixed sample) interaction (blue, grey is multiplied by a factor of 10 for clarity). The vertical error bars for the data points on this scale are smaller than the marker size.

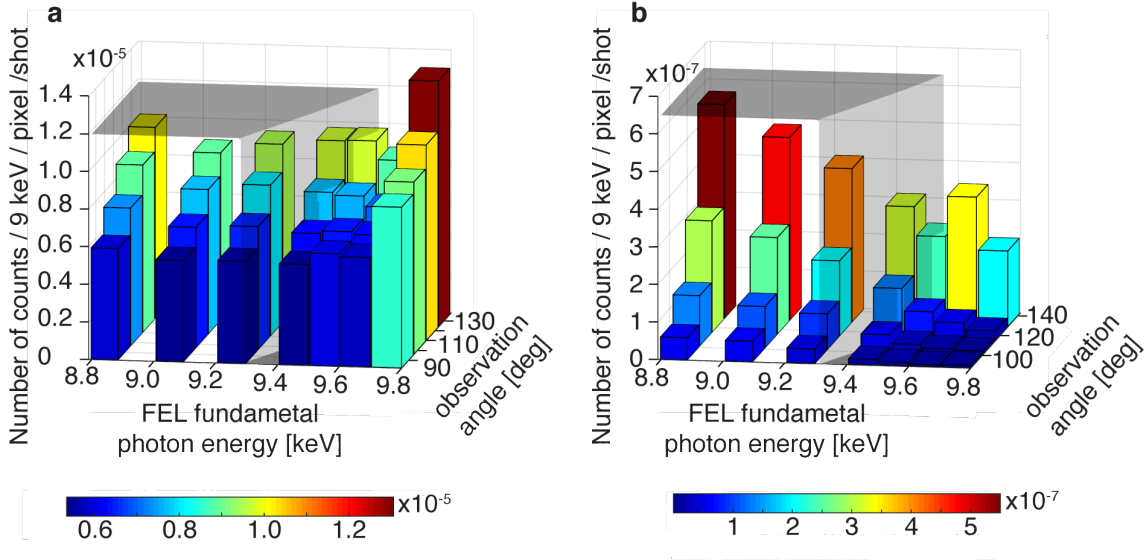

**Figure 4 | Photon energy scan.** The figure shows the angular-resolved observed scattering signal integrated around a photon energy of 18 keV for 100% FEL transmission with FEL fundamental photon energies ranging from 8.84 - 9.75 keV for the high-intensity interaction region **a)** and the low-intensity interaction region **b)**. The transparent grey curve shows where the scattered $n=2$ photon energy $\omega_2'(\omega_0, \theta)$ [$\omega_1'(2\omega_0, \theta)$] is equal to the Zr K-edge for **a** [**b**] according to eq. (1). The transmission contrast of the 250 μm thick Zr filters of $10^{-6}$ for photons just above (T= $1 \times 10^{-7}$) compared to just below (T = 10%) the K-edge (17.996 keV) was used for a coarse measure of the photon energy loss. The low-intensity signals show a cut-off for transmitted photon energies that increases with increasing observation angle, which is in agreement with $n=1$ linear Compton of the 2$^{nd}$ FEL harmonic. The high-intensity scattering signal does not show a decrease in magnitude even up to the highest measured photon energy of 9.75 keV. At this photon energy the expected Compton shift at 90º is ~717 eV (for both $n=2$ nonlinear scattering of 9.75 keV and linear scattering of 19.5 keV); therefore, the photon energy of the nonlinear signal $\omega_2'(\omega_0, \theta)$ has an additional redshift of at least 780 eV in order to be transmitted through the Zr foil. Note that the 9.75 keV curve is averaged only over 30k shots, whereas the 8.8 keV and 9.25 keV over

160k and all other curves over 60k shots). Note: the histograms from which fig. 4 is generated can be found in the supplementary information (SI).